# The Cosmic Ray Boron/Carbon Ratio Measured at Voyager and at AMS-2 from 10 MeV/nuc up to ~1 TeV/nuc and a Comparison With Propagation Calculations

W.R. Webber and T.L. Villa

New Mexico State University, Astronomy Department, Las Cruces, NM  88003, USA

## ABSTRACT

We have used new measurements of the B/C ratio in galactic cosmic rays at both low and high energies by the Voyager and AMS-2 spacecraft, respectively, along with propagation calculations using a truncated LBM to examine the implications of these new measurements over an extended energy range from a few MeV/nuc to ~1 TeV/nuc. We find that the predictions from both the truncated LBM and the Diffusive Reacceleration model for GALPROP both agree with the Voyager and AMS-2 measurements of the B/C ratio to within $\pm$ 10% throughout the entire energy range from ~50 MeV//nuc to ~1 TeV/nuc. The two propagation approaches also agree with each other to within $\pm$10% or less throughout this energy range. In effect a diffusion model, without significant additional acceleration, provides a match within $\pm$10% to the combined data from Voyager 1 and AMS-2 on the B/C ratio from ~50 MeV/nuc to ~1 TeV/nuc. The B/C ratio below ~50 MeV/nuc measured at V1 exceeds the predictions of both propagation models by as much as 3σ in the data measurement errors.

**Introduction**

The measurement of the Boron to Carbon ratio, B/C, in cosmic rays is of fundamental importance in understanding the origin and propagation of these cosmic rays in the galaxy to us. Carbon has a large abundance in the cosmic ray sources whereas Boron is believed to be totally absent. The amount of B produced depends on the cosmic ray propagation models which are described in terms of a spatial diffusion coefficient, D, and its dependence on rigidity, R. This will lead to a determination of the amount of matter and its dependence on energy traversed by cosmic rays, which is also important for many other studies involving the production of positrons, antiprotons, etc., in the galaxy. This dependence is usually described as $D \sim \beta P^x$, where x is taken to be between -0.33 to -0.50 depending on the scaling of the interstellar turbulence. But other effects, such as diffusive reacceleration, galactic winds, etc., could modify this rigidity dependence considerably. In fact, it is already known that the most obvious feature of B/C ratio is a peak at ~1 GeV/nuc corresponding to a rigidity ~3.3 GV. This feature has been ascribed to all of the processes mentioned above, singularly or in combination.

Earlier data on the B/C ratio at both low and high energies has at least two severe limitations, (1) Solar modulation effects change this ratio considerably below a few GeV/nuc and are really not all that well known; (2) The measurements above ~30 GeV/nuc did not have high accuracy.

Voyager 1 has now sampled the unmodulated interstellar spectra of B, C and other nuclei with energies of ~1 GeV/nuc down to a few MeV/nuc for ~5 years now beyond the heliopause (Cummings, et al., 2016; Webber, et al., 2017). At the other end of the spectrum the AMS-2 experiment has measured, to a precision of a few percent, the spectra of B, C and other nuclei above a few GeV/nuc up to ~1 TeV/nuc where solar modulation effects are negligible (Aguilar, et al., 2016).

It is the purpose of this paper to utilize these new measurements spanning the range from a few MeV/nuc to ~1 TeV/nuc to study the B/C ratio and also, for comparison, the N/B ratio and the implications of the two ratios for the amount of matter transversed by cosmic rays and other features related to the acceleration and propagation of cosmic rays.

**The Data**

The data on the B/C ratio from V1 below about 1.5 GeV/nuc (Cummings, et al., 2016; Webber, et al., 2017) and AMS-2 above ~3 GeV/nuc (Aguilar, et al., 2016) are shown in Figure 1. The calculations shown in this figure are from a truncated LBM described in Webber, et al., 2017, and the GALPROP-DR model described in Cummings, et al., 2016.

In Figure 2 we show another cosmic ray ratio of considerable interest, the N/B ratio. Nitrogen (N) is a secondary component of cosmic rays produced in the same manner as B. This secondary abundance dominates the observed total abundance of N at energies between ~100 MeV/nuc a few GeV/nuc. But $^{14}$N also has a primary or source component of abundance between 4-6% of $^{16}$O (e.g., Krombel and Wiedenbeck, 1988). This component dominates at both lower and higher energies where the interstellar path length becomes small thus reducing the production of secondary $^{15}$N. The data and calculations for the N/B ratio in Figure 2 are from the same references and calculations as for B/C. Included for the N/B ratio are the measurements from ACE spacecraft in 2009 when the solar modulation was at a low level ~300 MV and also HEAO-3 observations between 1 GeV/nuc and 35 GeV/nuc (Lave, et al., 2013; Engelmann, et al., 1990).

**Comparison of the Results and the Calculations from Galactic Propagation Models**

For all LBM calculations here we have assumed source spectra of the form dj/dP ~P$^{-2.24}$, which are the same for N and C. As a result, one would expect the predicted B/C intensity ratios and their agreement or non-agreement with the measurements would be all about the interstellar propagation itself.

**a) Fitting the B/C Ratio**

For the situation illustrated in Figure 1, above about 30-50 MeV/nuc, both the truncated LBM and the GALPROP-DR models agree with each other to within ±10% over the entire energy range up to ~1 TeV/nuc where the calculated B/C ratio is between 4-5%, the same as that measured by AMS-2. Even in the energy range from 100 to 1000 MeV/nuc where the DR version has usually been employed to better reproduce the intermediate energy spectra (and where there is also a lack of definitive low solar modulation data), the two propagation approaches do not predict a significant difference in the B/C ratio. Overall, both models agree with the Voyager data at energies ~1 GeV/nuc and below and the AMS-2 data from 3 GeV/nuc to 1 TeV/nuc to within $\pm$10% or better. This comparison is shown in Table II for the AMS-2 average data and calculations using the truncated LBM and GALPROP-DR models.

For the truncated LBM used here, (originating in Engelmann, et al., 1990, and used by Webber and Higbie, 2009, 2015, and Webber, et al., 2017), the mean path length $\lambda$ is taken to be = 18.2 $\beta$ $P^{-0.45}$ above P=1.0 GV (=125 MeV/nuc for A/Z=2.0 nuclei), where the coefficients of 18.2 and -0.45 are determined from the measured unmodulated B/C ratio at both low and high energies, so, in fact, the LBM should fit the data. The value of -0.45 for the exponent of P corresponds to D ~$P^{0.45}$ where D is the isotropic diffusion coefficient. In this model the average density is 0.4 g/cm$^2$ of 90% H and 10% He and the value of the diffusion coefficient is selected so that the "age" of cosmic rays at ~ 2 GV is 1.5 x 10$^7$ year in accordance with the measurements of $^{10}$Be (Yanasak, et al., 2001). Below ~1 GV the path length =9 g/cm$^2$ in this model.

In the GALPROP-DR model compared here, the situation is more complicated since the diffusive reacceleration used in the model may modify the rigidity dependence of the B/C ratio obtained from the calculation and also in this model, D itself, is taken to be ~$P^{-0.33}$ in accordance with a Kolmogorov turbulence spectrum.

The interesting thing is that both calculations give nearly the same B/C ratio within $\pm$10%, both above and below the peak in this ratio which is at about 1 GeV/nuc. The idea that reacceleration is needed to predict the correct B/C ratio at energies just below the peak (e.g., Drury and Strong, 2017) is not supported by the calculations that are compared in this paper.

The AMS-2 observations of the B/C ratio above a few GeV/nuc are, on average, a good match for what is obtained for diffusive propagation in a LBM with D ~$\beta$ $P^{-0.45}$. The situation below 30-50 MeV/nuc, where there is an excess of observed B over both of the propagation calculations will be discussed later.

The truncated LBM calculations are made with a path length $\lambda$=18.2 $\beta$ $P^{-0.45}$ above ~1 .0 GV. This path length in terms of g/cm$^2$ is shown in Figure 3. At 100 GeV/nuc this mean path length is ~1.5 g/cm$^2$ and at 1 TeV/nuc it is still ~0.5 g/cm$^2$. These path lengths are appropriate for calculating the secondary positron and anit-proton production in corresponding propagation models in the galaxy.

Table III shows the calculated B, C and N intensities from 1 MeV/nuc to 1 TeV/nuc for a truncated LBM with the path length $\lambda = 18.2 \ \beta \ P^{-0.45}$ and $\lambda = 9 \ g/cm^2$ below ~1 GV.

**Fitting the N/B Intensity Ratio**

This ratio, shown in Figure 2, brings into play not only the propagation history, which should be the same for both B and N nuclei, but also the source component for N which has been estimated to be 4-6% for the $^{14}N/^{16}O$ ratio. The propagation component exceeds the source component between a few hundred MeV/nuc and a few GeV/nuc where the path length is larger than a few $g/cm^2$. At both low and high energies, however, the N source component begins to dominate and since B has no source component the N/B ratio rapidly increases. The data, which includes only HEAO data above ~1 GeV/nuc up to ~35 GeV/nuc, illustrates this effect very well with a minimum N/B ratio of ~0.9 between a few hundred MeV/nuc and a few GeV/nuc which increases to a ratio to ~1.5 at the highest energies. At energies less than 100 MeV/nuc the ratio also increases to ~1.3.

The predictions of both the LBM and GALPROP model agree with the measurements at energies above ~1 GeV/nuc. Future AMS-2 measurements of N at the highest energies would help to define the N source abundance which, in fact, determines how rapidly this ratio increases above a few GeV/nuc. At lower energies the GALPROP-DR calculations exceed the truncated LBM calculations and also the measurements at Voyager (Cummings, et al., 2016) and ACE (Lave, et al., 2013) by at least 20-30%, well outside the experimental error. The truncated LBM with a source abundance of N/O = 5% +1% agrees with the Voyager measurements.

**The B/C Ratio Below ~50 MeV/nuc**

It is seen from Figure 1 that the B/C ratio measured at Voyager below ~50 MeV/nuc is essentially constant at a value of about 0.13. The statistical errors are +15-20% on the lowest two energy points. This alone could account for at least 0.3 of the difference between predictions and measurements. This measured ratio could also be achieved in a LBM propagation picture by an increasing path length at energies below ~30 MeV/nuc reaching at least 15 g/cm$^2$ at 10-20 MeV/nuc. It could also be achieved if the cross sections into B at low energies, e.g., mainly $^{12}$C→$^{11}$B, $^{16}$O→$^{11}$B, etc., are underestimated by as much as 20-30%. These cross sections below ~100 MeV/nuc are not well measured.

This flat B/C ratio could also be achieved if there were a "source" component for B such as might occur if the cosmic rays initially passed through 0.1-0.2 g/cm$^2$ of matter near the source (D'Angelo, Blasi and Amato, 2016) as in a nested LBM (Cowsik and Wilson, 1975).

This NLBM scenario would also modify the "source" abundance for N (and other mostly secondary components) so that the apparent "low" energy N source abundance would be less than the apparent "high" energy N source abundance.

This scenario, with the amount of material at the source independent of energy, would also produce an excess of B at high energies which would make the high energy predictions now larger than the AMS-2 observations. Hopefully further measurements of the low energies for B and N nuclei will help to clarify the situation.

## Summary and Conclusions

We have used new measurements of the B/C ratio at low and high energies by the Voyager and AMS-2 spacecraft, along with propagation calculations using a truncated LBM to examine the implications of these new measurements over an extended energy range from a few MeV/nuc to ~1 TeV/nuc. We find that the predictions from both the truncated LBM and the GALPROP Diffusive Reacceleration model agree with the Voyager and AMS-2 measurements of the B/C ratio to within $\pm$ 10% throughout the entire energy range from ~50 MeV//nuc to ~1 TeV/nuc. The two propagation approaches also agree with each other to within $\pm$10% or less throughout this energy range including the energies just below the peak in the B/C ratio around 1 GeV/nuc.

For the N/B ratio also examined here, the truncated LBM also matches the measured N/B ratios within ~$\pm$10% over the entire energy range covered by the Voyager and HEAO-3 measurements from 10 MeV/nuc to ~35 GeV/nuc for an $^{14}$N/O source abundance of 5 $\pm$ 1%, recognizing the level of uncertainty of the lowest energy measurements and the cross sections for B production.

The GALPROP-DR diffusion reacceleration model compared here considerably over-predicts the measured N/B ratio at energies below ~100 MeV/nuc.

Concerning the significance of the differences between the Voyager measurements of the B/C ratio at energies below ~50 MeV/nuc and the predictions of both the truncated LBM and the GALPROP-DR model one should again recognize the level of uncertainty in both the measured Boron intensity values and the cross sections for B production at these lowest energies.

**<u>Acknowledgements:</u>** The authors are grateful to the Voyager team that designed and built the CRS experiment with the hope that one day it would measure the galactic spectra of nuclei and electrons. This includes the present team with Ed Stone as PI, Alan Cummings, Nand Lal and Bryant Heikkila, and to others who are no longer members of the team, F.B. McDonald and R.E. Vogt. Their prescience will not be forgotten. This work has been supported throughout the more than 40 years since the launch of the Voyagers by the JPL.

| TABLE I |
| :---: |
| **Calculated and Measured** |
| **B/C Ratio in Percent at High Energies** |

| Energy (GeV/nuc) | Truncated LBM | Measured B/C (AMS-2) | GALPROP |
| :---: | :---: | :---: | :---: |
| 3.16 | 26.0 | 27.2 | 28.5 |
| 10 | 20.1 | 20.5 | 20.2 |
| 31.6 | 14.3 | 14.5 | 14.1 |
| 100 | 9.6 | 9.6 | 10.0 |
| 316 | 6.4 | $6.6\pm0.6$ | 7.3 |
| 1000 | 4.2 | $4.5\pm0.5$ | 5.2 |

| TABLE II |
|---|
| **TRUNCATED LBM CALCULATION WITH** |
| $\lambda = 18.2 \, \beta \, P^{-0.45}$; $= 9 g/cm^2$ above 1.0 GV |
| **INTENSITES P/m²·sr·s·MeV/nuc** |

| Energy | Sum B | Sum C | Sum N |
|---|---|---|---|
| 7.5 | 1.33E-03 | 1.69E-02 | 1.80E-03 |
| 10 | 1.69E-03 | 2.08E-02 | 2.23E-03 |
| 13 | 2.12E-03 | 2.47E-02 | 2.70E-03 |
| 18 | 2.65E-03 | 2.88E-02 | 3.24E-03 |
| 24 | 3.26E-03 | 3.29E-02 | 3.82E-03 |
| 32 | 3.94E-03 | 3.65E-02 | 4.42E-03 |
| 42 | 4.67E-03 | 3.93E-02 | 5.01E-03 |
| 56 | 5.36E-03 | 4.04E-02 | 5.51E-03 |
| 75 | 5.94E-03 | 4.01E-02 | 5.88E-03 |
| 100 | 6.29E-03 | 3.84E-02 | 6.08E-03 |
| 133 | 5.95E-03 | 3.49E-02 | 5.80E-03 |
| 178 | 5.80E-03 | 3.02E-02 | 5.54E-03 |
| 237 | 5.34E-03 | 2.49E-02 | 5.03E-03 |
| 316 | 4.64E-03 | 1.94E-02 | 4.30E-03 |
| 422 | 3.74E-03 | 1.43E-02 | 3.43E-03 |
| 562 | 2.80E-03 | 1.01E-02 | 2.56E-03 |
| 750 | 1.96E-03 | 6.78E-03 | 1.80E-03 |
| 1000 | 1.30E-03 | 4.40E-03 | 1.21E-03 |
| 1334 | 7.98E-04 | 2.73E-03 | 7.54E-04 |
| 1779 | 4.65E-04 | 1.64E-03 | 4.50E-04 |
| 2372 | 2.58E-04 | 9.48E-04 | 2.55E-04 |
| 3163 | 1.39E-04 | 5.35E-04 | 1.41E-04 |
| 4218 | 7.18E-05 | 2.93E-04 | 7.55E-05 |
| 5624 | 3.59E-05 | 1.56E-04 | 3.91E-05 |
| 7500 | 1.75E-05 | 8.12E-05 | 1.97E-05 |
| 10001 | 8.30E-06 | 4.14E-05 | 9.66E-06 |
| 13337 | 3.84E-06 | 2.07E-05 | 4.64E-06 |
| 17785 | 1.74E-06 | 1.02E-05 | 2.18E-06 |
| 23717 | 7.75E-07 | 4.96E-06 | 1.01E-06 |
| 31627 | 3.40E-07 | 2.39E-06 | 4.63E-07 |
| 42176 | 1.47E-07 | 1.14E-06 | 2.10E-07 |
| 56242 | 6.33E-08 | 5.38E-07 | 9.40E-08 |
| 75000 | 2.69E-08 | 2.53E-07 | 4.19E-08 |
| 100014 | 1.14E-08 | 1.18E-07 | 1.86E-08 |
| 133371 | 4.78E-09 | 5.49E-08 | 8.18E-09 |
| 177853 | 1.99E-09 | 2.54E-08 | 3.60E-09 |
| 237171 | 8.28E-10 | 1.17E-08 | 1.57E-09 |
| 316273 | 3.42E-10 | 5.37E-09 | 6.88E-10 |
| 421757 | 1.41E-10 | 2.45E-09 | 3.00E-10 |
| 562422 | 5.78E-11 | 1.12E-09 | 1.30E-10 |
| 750001 | 2.36E-11 | 5.05E-10 | 5.65E-11 |
| 1000143 | 9.58E-12 | 2.27E-10 | 2.44E-11 |

## REFERENCES

Aguilar, M., Cavasonza, L.A., Ambrosi, G., et al., 2016, Phys. Rev. Lett., 117, 231102

Cowsik, R. and Wilson, L.W., 1975, Proc. 14[th] ICRC (Munich), 2, 659

Cummings, A.C., Stone, E.C., Heikkila, B.C., et al., 2016, Ap.J., 831, 21

D'Angelo, M., Blasi, B. and Amato, E, 2016, PhRvD, 94, 3003

Drury, L.O. and Strong, A.W., 2017, A&A, 597, A117

Engelmann, J.J., Ferrando, P., Soutoul, A., et al., 1990, A&A, 233, 96

Krombel, K.E. and Wiedenbeck, M.E., 1988, Ap.J., 328, 940

Lave, K., Wiedenbeck, M.E., Binns, W.R., et al., 2013, Ap.J., 730, 117

Webber, W.R., et al., 2017, (to be submitted to arXiv)

Webber, W.R. and Higbie, P.R., 2009, JGR, 114, 2103

Webber, W.R. and Higbie, P.R., 2015, http://arXiv.org/abs/1503.05891

Yanasuk, N.E, Wiedenbeck, M.E., Mewaldt, R.A., et al., 2001, Ap.J., 563, 768-792

## FIGURE CAPTIONS

**Figure 1:** The B/C ratio as a function of energy. The measurements of this ratio below ~2.4 GeV/nuc are from Voyager 1 beyond the heliopause (Cummings, et al., 2016; Webber, et al., 2017). The higher energy measurements are from AMS-2 (Aguilar, et al., 2016). The calculated ratios are from the truncated LBM model described in the text and the GALPROP-DR model (Cummings, et al., 2016).

**Figure 2:** The N/B ratio as a function of energy. The measurements at lower energies are from V1 (Cummings, et al., 2016) and ACE near the Earth (Lave, et al., 2013). The measurements at higher energies are from HEAO-3 (Engelmann, et al., 1990). The ratio calculated for a modulation potential = 300 MV is shown as a dashed line.

**Figure 3:** The mean path length as a function of rigidity and energy/nuc. This is determined by $\lambda=18.2 \beta P^{-0.45}$ in $g/cm^2$. Mean path lengths for different values of $P_0$, with $\lambda \sim \beta$ below $P_0$ and for a constant path length = 9 $g/cm^2$ below 1 GV are shown. Also the energy/nuc for A/Z = 2 nuclei is shown along the bottom axis.

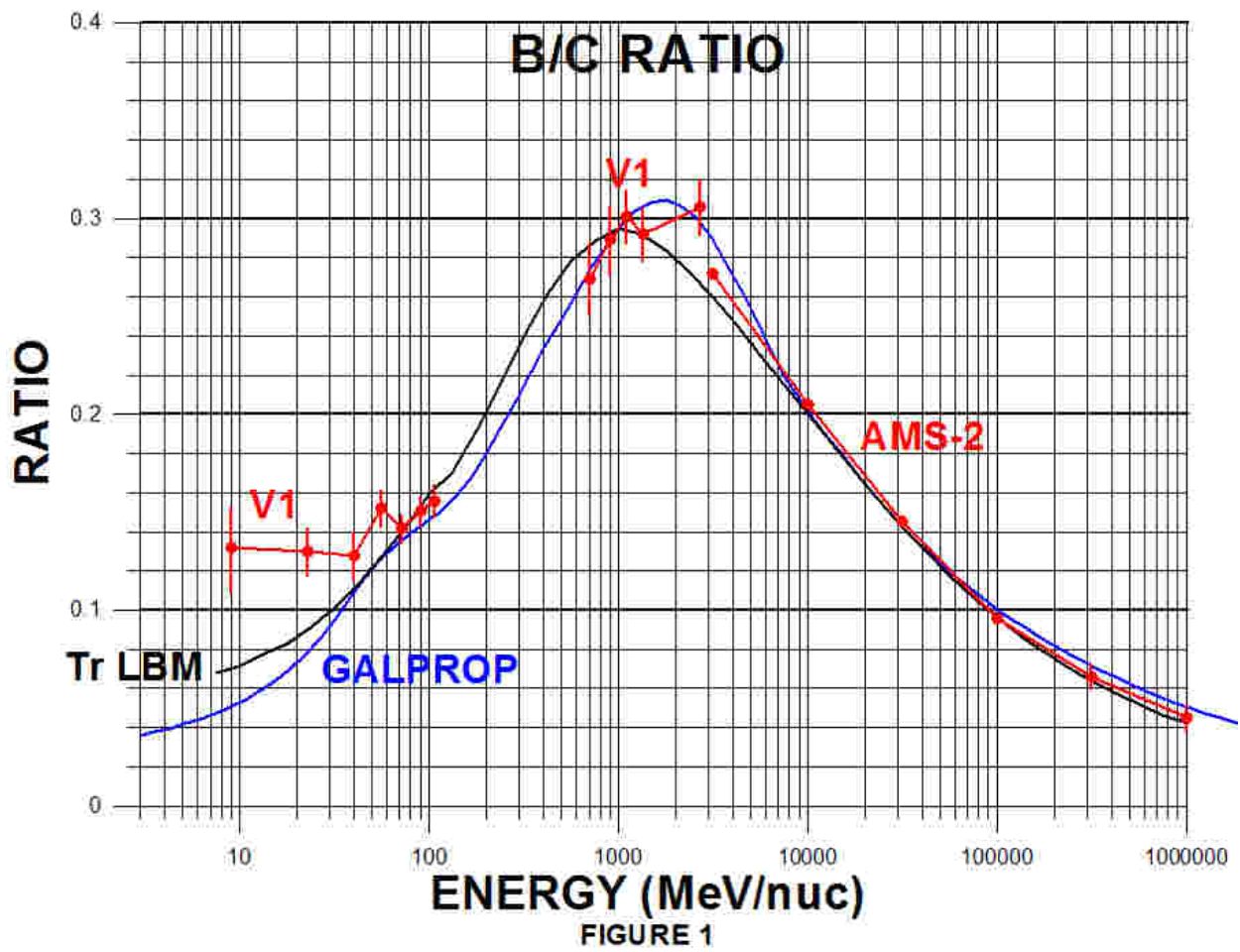

FIGURE 1

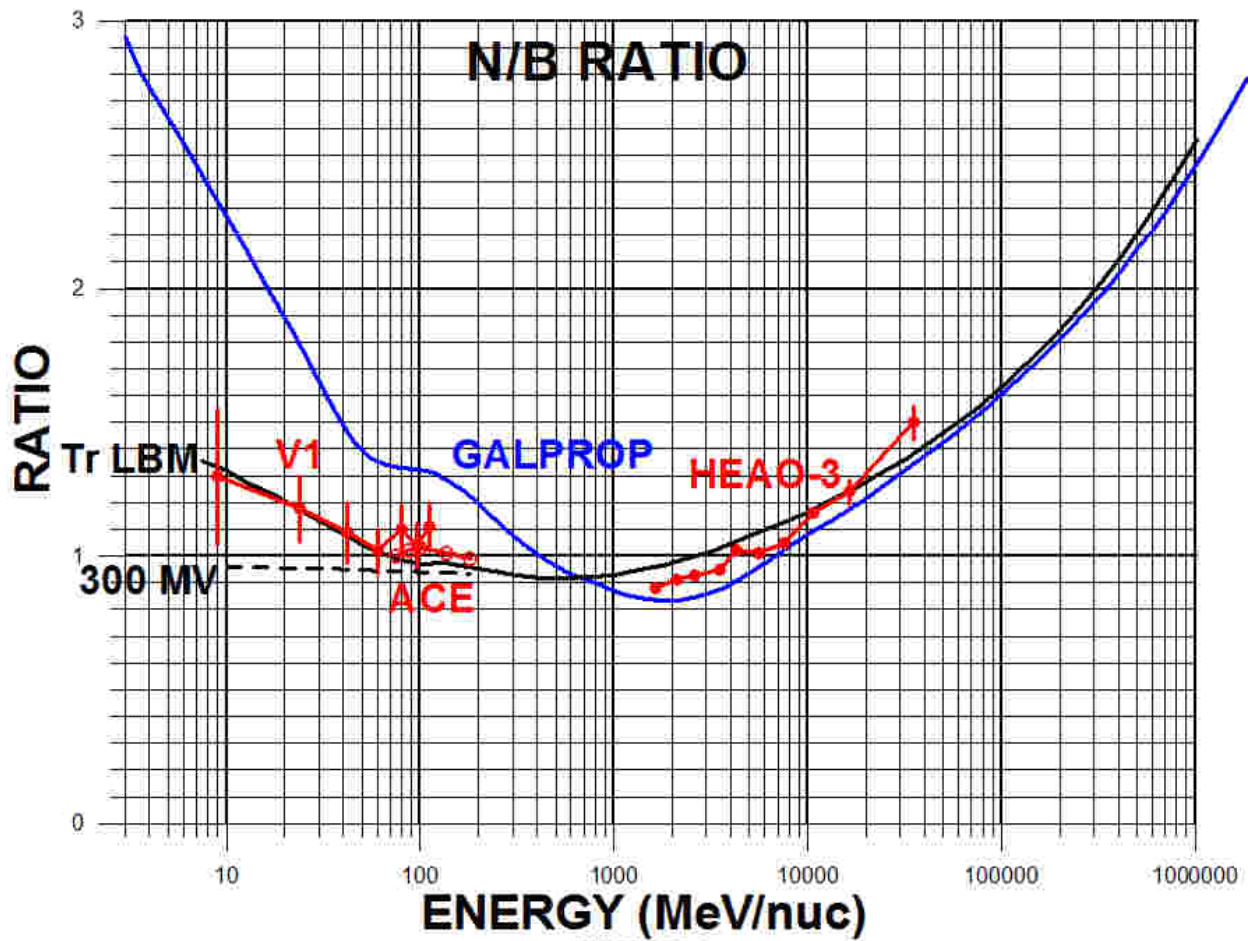

FIGURE 2

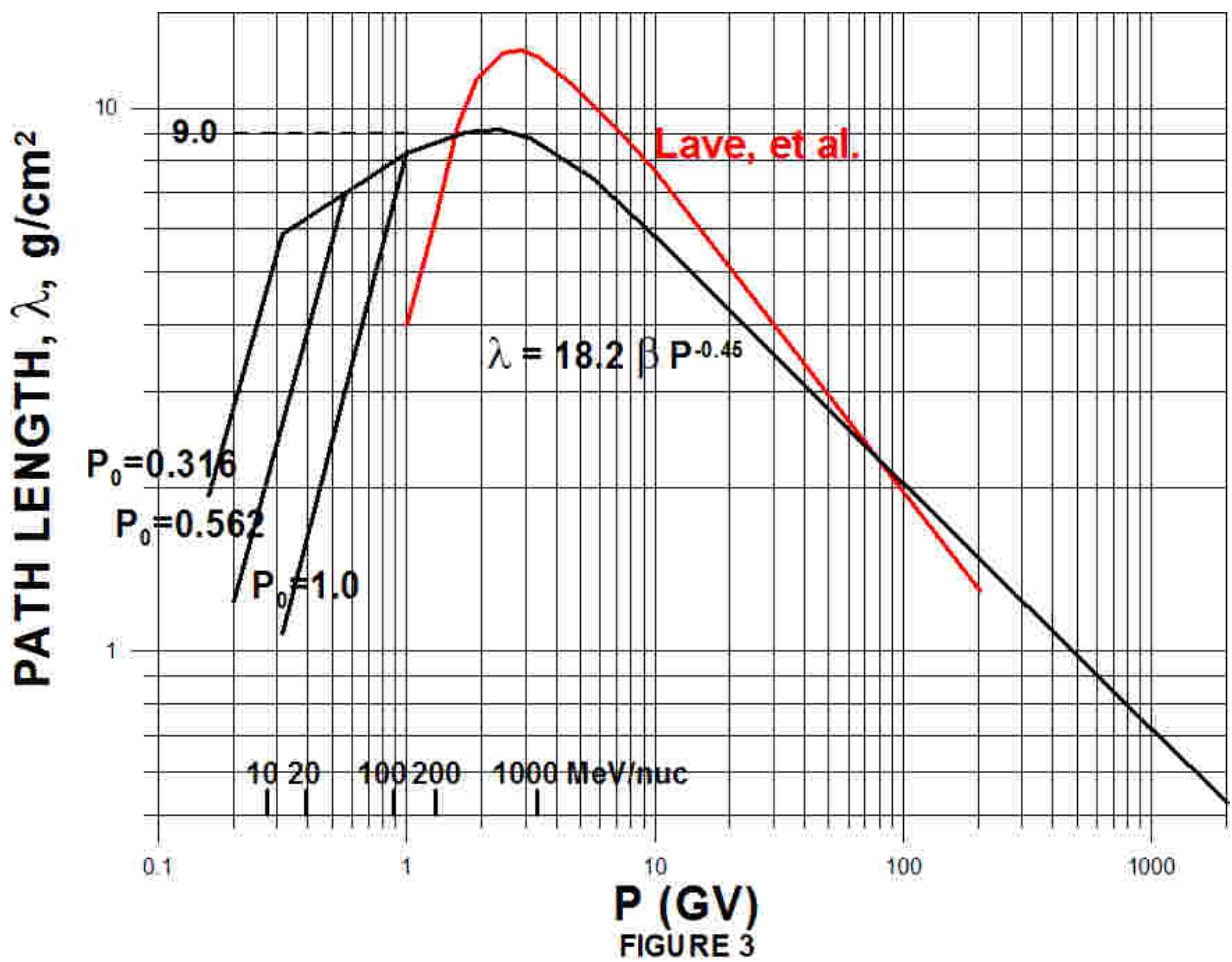

FIGURE 3